\documentclass[a4paper,12pt]{article}
\usepackage{epsfig}
\usepackage{amsmath}
\usepackage{amssymb}
\usepackage{amsfonts}
\usepackage{bbm}
\usepackage{fleqn}
\usepackage[footnotesize]{caption}
\usepackage{graphicx}
\usepackage{mathrsfs}
\usepackage[center,footnotesize,hang]{subfigure}

\def\I{\mathrm{i}}

\def\be{\begin{equation}}
\def\ee{\end{equation}}
\def\beq{\begin{equation}}
\def\eeq{\end{equation}}
\def\bea{\begin{eqnarray}}
\def\eea{\end{eqnarray}}

\def\<{\left\langle}
\def\>{\right\rangle}

\allowdisplaybreaks[2]
\begin{document}
\bibliographystyle{OurBibTeX}
\begin{titlepage}
 \vspace*{-15mm}
\begin{flushright}
\end{flushright}
\vspace*{5mm}
\begin{center}
{ \sffamily \LARGE Parametrizing the lepton mixing matrix in terms of deviations from
tri-bimaximal mixing}
\\[8mm]
S.~F.~King\footnote{E-mail:
\texttt{sfk@hep.phys.soton.ac.uk}}
\\[3mm]
{\small\it
School of Physics and Astronomy,
University of Southampton,\\
Southampton, SO17 1BJ, U.K.
}\\[1mm]
\end{center}
\vspace*{0.75cm}
\begin{abstract}
\noindent

We propose a parametrization of the lepton mixing matrix in terms of
an expansion in powers of the deviations of the reactor, solar and
atmospheric mixing angles from their tri-bimaximal values.  We show
that unitarity triangles and neutrino oscillation formulae have a very
compact form when expressed in this parametrization, resulting in
considerable simplifications when dealing with neutrino
phenomenology. The parametrization, which is completely general,
should help to establish possible relations between the deviations of
the reactor, solar and atmospheric mixing angles from their
tri-bimaximal values, and hence enable models which predict such
relations to be more directly compared to experiment.

\end{abstract}
\end{titlepage}
\newpage
\setcounter{footnote}{0}
Over the last decade neutrino physics has undergone a revolution
with the measurement of neutrino mass and lepton mixing from
a variety of solar, atmospheric and terrestrial neutrino
oscillation experiments \cite{Bahcall:2004cc}.
Lepton mixing is described by the $3\times 3$ matrix
\cite{MNS}
\begin{equation}
U =\left(\begin{array}{ccc}
U_{e1} & U_{e2} & U_{e3} \\
U_{\mu1} & U_{\mu2} & U_{\mu3} \\
U_{\tau1} & U_{\tau2} & U_{\tau3} \\
\end{array}\right).
\label{MNS}
\end{equation}
The Particle Data Group (PDG) parameterization of the
lepton mixing matrix (see e.g.\ \cite{PDG}) is:
\begin{eqnarray}\label{Eq:StandardParametrization}
 U = \left(
  \begin{array}{ccc}
  c_{12}c_{13} &
  s_{12}c_{13} & s_{13}e^{-\I\delta}\\
  -c_{23}s_{12}-s_{13}s_{23}c_{12}e^{\I\delta} &
  c_{23}c_{12}-s_{13}s_{23}s_{12}e^{\I\delta}  &
  s_{23}c_{13}\\
  s_{23}s_{12}-s_{13}c_{23}c_{12}e^{\I\delta} &
  -s_{23}c_{12}-s_{13}c_{23}s_{12}e^{\I\delta} &
  c_{23}c_{13}
  \end{array}
  \right) \, P \, ,
\end{eqnarray}
where $s_{13}=\sin \theta_{13}$, $c_{13}=\cos \theta_{13}$
with $\theta_{13}$ being the
reactor angle, $s_{12}=\sin \theta_{12}$, $c_{12}=\cos \theta_{12}$
with $\theta_{12}$ being the
solar angle, $s_{23}=\sin \theta_{23}$, $c_{23}=\cos \theta_{23}$
with $\theta_{23}$ being the
atmospheric angle, $\delta$ is the (Dirac) CP violating phase which is in
principle measurable in neutrino oscillation experiments, and
$P = \mathrm{diag}(e^{\I \tfrac{\alpha_1}{2}},
e^{\I\tfrac{\alpha_2}{2}}, 0)$ contains additional (Majorana)
CP violating phases $\alpha_1, \alpha_2$.
Current data is consistent with the tri-bimaximal mixing (TBM) form
\footnote{Sometimes an alternative phase convention is chosen in which the
third row of $U_{\mathrm{MNS}}$ has its signs reversed.}
\cite{HPS}
\begin{eqnarray}
U \approx
\left( \begin{array}{rrr}
\sqrt{\frac{2}{3}}  & \frac{1}{\sqrt{3}} & 0 \\
-\frac{1}{\sqrt{6}}  & \frac{1}{\sqrt{3}} & \frac{1}{\sqrt{2}} \\
\frac{1}{\sqrt{6}}  & -\frac{1}{\sqrt{3}} & \frac{1}{\sqrt{2}}
\end{array}
\right)P.
\label{MNS0}
\end{eqnarray}
Many models can account for TBM lepton mixing
\cite{King:2005bj,Frampton:2004ud,Altarelli:2006kg,Ma:2007wu,deMedeirosVarzielas:2005ax,Harrison:2003aw,Chan:2007ng}.
However there is no convincing reason for TBM to be exact,
and in the future deviations from it are expected to be observed.
With this in mind it is clearly useful to develop a parametrization
of the lepton mixing matrix in which such deviations are manifest,
and in which the predictions of models for deviations from tri-bimaximal
mixing can naturally be expressed. Such a parametrization must
be model independent, and completely general so that it can be used
by experimentalists and phenomenologists in performing analyses of neutrino
experiments. It must also be sufficiently simple to be useful and yet accurate
enough to be reliable.

In this paper we discuss a parametrization of the lepton mixing matrix
which possesses all of the above desirable features.
The parametrization exploits the empirical observed closeness of lepton mixing to
the TBM form, and is analagous to the Wolfenstein parametrization of quark mixing
\cite{Wolfenstein:1983yz}.
Just as the Wolfenstein parametrization is an expansion about the unit
matrix, so the present parametrization is an expansion about the tri-bimaximal matrix.
Unlike the Wolfenstein parametrization, we introduce three small parameters
parametrizing the deviations of the
reactor, solar and atmospheric angles from their tri-bimaximal values.
The expansion works since all three parameters are empirically
small, having magnitude of order the Wolfenstein parameter
$\lambda \approx 0.227$ or less.
A related proposal to expand the lepton mixing matrix elements
about the tri-bimaximal matrix elements, using a different
parametrization from that introduced here, was discussed in
\cite{Li:2004dn}. 
Other related proposals to parametrize the lepton mixing matrix
have been considered in
\cite{Parke,Xing:2002sw,Bjorken:2005rm,Paul,Datta:2005ci}.\footnote{I am grateful to S.~Parke, Z.Z.~Zing and P.~Harrison for informing
me about their work \cite{Parke,Xing:2002sw,Paul}.}

Without loss of generality we define
\be
s_{13} = \frac{r}{\sqrt{2}}, \ \ s_{12} = \frac{1}{\sqrt{3}}(1+s),
\ \ s_{23} = \frac{1}{\sqrt{2}}(1+a),
\label{rsa}
\ee
where we have introduced the three real parameters
$r,s,a$ to describe the deviations of the reactor, solar and
atmospheric angles from their tri-bimaximal values.
Global fits of the conventional mixing angles
\cite{Maltoni:2004ei} can be translated into the $2\sigma$ ranges
\footnote{Note that $r$ must be positive definite,
while $s,a$ can take either sign. Indeed there is a preference for $s$ to be negative.}
\be
0<r<0.22, \ -0.11<s<0.04, \ -0.12<a<0.13.
\ee
The empirical smallness of these parameters suggests that we
consider an expansion of the lepton mixing matrix in powers of $r,s,a$
about the tri-bimaximal form.
To first order
\footnote{The second order corrections are expected
to be very small, of order one per cent or less,
depending on the (presently constrained but undetermined) values of $r,s,a$.
Throughout the main text where results are presented to first
order in $r,s,a$, the second order corrections are given in Appendix \ref{second}.
}
in $r,s,a$ the lepton mixing matrix can be written
\begin{eqnarray}
U \approx
\left( \begin{array}{ccc}
\sqrt{\frac{2}{3}}(1-\frac{1}{2}s)  & \frac{1}{\sqrt{3}}(1+s) & \frac{1}{\sqrt{2}}re^{-i\delta } \\
-\frac{1}{\sqrt{6}}(1+s-a + re^{i\delta })  & \frac{1}{\sqrt{3}}(1-\frac{1}{2}s-a- \frac{1}{2}re^{i\delta })
& \frac{1}{\sqrt{2}}(1+a) \\
\frac{1}{\sqrt{6}}(1+s+a- re^{i\delta })  & -\frac{1}{\sqrt{3}}(1-\frac{1}{2}s+a+ \frac{1}{2}re^{i\delta })
 & \frac{1}{\sqrt{2}}(1-a)
\end{array}
\right)P.
\label{MNS1}
\end{eqnarray}
As in the Wolfenstein parametrization, the above parametrization of the lepton
mixing matrix avoids the introduction of mixing angles,
instead dealing directly with elements of the mixing matrix.
Accordingly the parametrization results in
considerable simplifications when dealing with neutrino phenomenology.
For example, the complex elements of the quark mixing matrix can be visualized
using unitarity triangles \cite{Chau:1984fp}, which, when normalized,
only depend on two parameters.
The same proves to be true when using the above
parametrization of the lepton mixing matrix.
The sides of the unitarity triangles enter into
the neutrino oscillation formulae, and consequently
these are also considerably simplified by the new parametrization.
In the remainder of the paper we shall discuss
unitarity triangles and neutrino oscillation formulae
using the above parametrization.

CP violation is described by the Jarlskog \cite{Jarlskog:2006za}
invariant which to leading order is
\be
J\approx \frac{r}{6}\sin \delta .
\label{J}
\ee
Leptonic unitarity triangles \cite{Fritzsch:1999ee}
may be constructed using the orthogonality
of different pairs of columns or rows of the mixing matrix. Only the opening angles,
side lengths and areas of the triangles have physical significance.
For example the area of each unitarity triangle is $A=\frac{1}{2}|J|$ and
CP violation implies that the longest side of each unitarity triangle is
smaller than the sum of the other two. Current solar, reactor and atmospheric
experiments directly constrain the elements $U_{e2}$, $U_{e3}$
and $U_{\mu 3}$, which have a particularly simple
parametrization in Eq.\ref{MNS1}. The most important unitarity triangles
should therefore include all of the elements $U_{e2}$, $U_{e3}$
and $U_{\mu 3}$. There are two such unitarity triangles, the $\nu_2.\nu_3$
one \cite{Bjorken:2005rm}
corresponding to the orthogonality of the second and third column,
and the $\nu_{e}.\nu_{\mu}$ one \cite{Farzan:2002ct}
corresponding to the orthogonality of the first
and second row. Each of them has a simple expression in terms of the new parametrization,
as we now discuss.

\begin{figure}[t]
\centering
\includegraphics[width=0.96\textwidth]{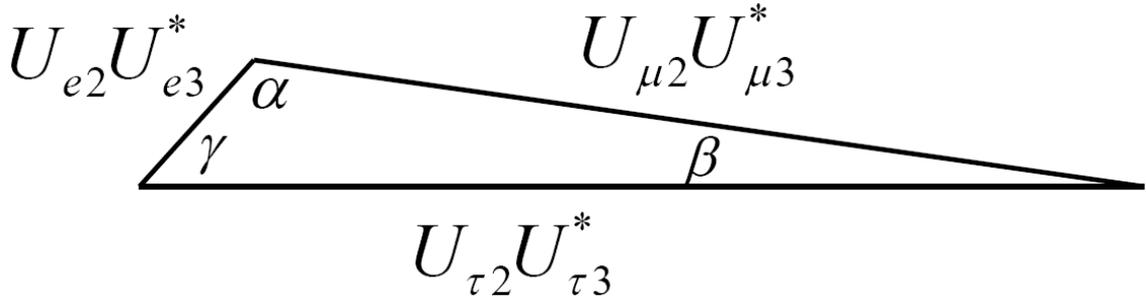}
\vspace*{-4mm}
    \caption{The $\nu_2.\nu_3$ unitarity triangle. The angle $\gamma$ is equal
    to the CP phase $\delta$ to first order. The unknown Majorana phases just rotate the triangle
    in the complex plane. The rescaled triangle is oriented as shown with the opening angles unchanged,
    the horizontal side having unit length, and the shortest side having length $r$ to first order.
    Currently $0<r<0.22$ at $2\sigma$, and the opening angles $\alpha$, $\beta$ and $\gamma$ are all undetermined.
} \label{UT1}
\vspace*{-2mm}
\end{figure}

The $\nu_2.\nu_3$ triangle in Fig.\ref{UT1} corresponds to the unitarity relation
\be
U_{e2}U_{e3}^*+U_{\mu 2}U_{\mu 3}^*+U_{\tau 2}U_{\tau 3}^*=0.
\label{23}
\ee
To first order the sides of this unitarity triangle
are given by
\bea
S_1 & = & U_{e2}U_{e3}^*\approx \frac{1}{\sqrt{6}}re^{i\delta} \nonumber \\
S_2 & = & U_{\mu 2}U_{\mu 3}^*\approx \frac{1}{\sqrt{6}}(1-\frac{s}{2}-\frac{r}{2}e^{i\delta} )\nonumber \\
S_3 & = & U_{\tau 2}U_{\tau 3}^*\approx -\frac{1}{\sqrt{6}}(1-\frac{s}{2}+\frac{r}{2}e^{i\delta} ).
\label{S}
\eea
Clearly $S_1+S_2+S_3=0$ to first order. The invariant J is
\be
J=Im(S_1S_2^*)=Im(S_3S_1^*)=Im(S_2S_3^*)
\ee
which yields Eq.\ref{J}. To first order the sides of
this triangle are only sensitive to the solar and reactor parameters
$s$ and $r$ and the phase $\delta$, with the atmospheric parameter $a$ only appearing at second order.
One may rescale the sides by $S_3$
\bea
S_1' & = & \frac{U_{e2}U_{e3}^*}{U_{\tau 2}U_{\tau 3}^*}\approx -re^{i\delta} \nonumber \\
S_2' & = & \frac{U_{\mu 2}U_{\mu 3}^*}{U_{\tau 2}U_{\tau 3}^*}\approx -1+re^{i\delta}\nonumber \\
S_3' & = & 1.
\label{Sp}
\eea
To first order the rescaled triangle is only sensitive to the reactor parameter $r$ and the phase $\delta$,
which is the anticipated result. To second order
the solar parameter $s$ (but not the atmospheric parameter $a$) appears.

\begin{figure}[t]
\centering
\includegraphics[width=0.96\textwidth]{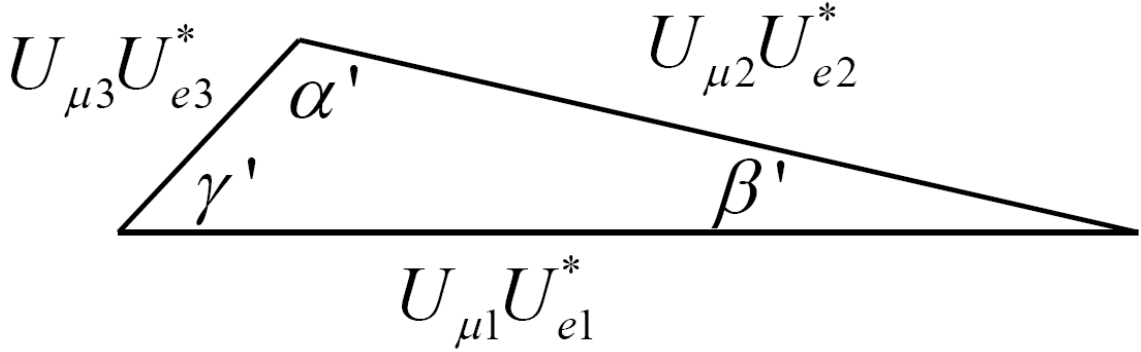}
\vspace*{-4mm}
    \caption{The $\nu_e.\nu_{\mu}$ unitarity triangle. The angle $\gamma'$ is equal
    to the CP phase $\delta$ to first order. The unknown Majorana phases cancel.
    The rescaled triangle is oriented as shown with the opening angles unchanged,
    the horizontal side having unit length, and the shortest side having
    length $\frac{3}{2}r$ to first order. Currently $0<r<0.22$ at $2\sigma$
    and the opening angles $\alpha'$, $\beta'$ and $\gamma'$ are all undetermined.
} \label{UT2}
\vspace*{-2mm}
\end{figure}

The other unitarity triangle of interest is $\nu_e.\nu_{\mu}$
in Fig.\ref{UT2} corresponding to the unitarity relation
\be
U_{\mu 1}U_{e1}^*+U_{\mu 2}U_{e2}^*+U_{\mu 3}U_{e 3}^*=0.
\label{23}
\ee
To first order the sides of this unitarity triangle
are given by
\bea
T_1 & = & U_{\mu 1}U_{e1}^*\approx -\frac{1}{3}(1+\frac{s}{2}-a+re^{i\delta} )\nonumber \\
T_2 & = & U_{\mu 2}U_{e2}^*\approx \frac{1}{3}(1+\frac{s}{2}-a-\frac{r}{2}e^{i\delta} )\nonumber \\
T_3 & = & U_{\mu 3}U_{e3}^*\approx \frac{1}{2}re^{i\delta}.
\label{T}
\eea
Clearly $T_1+T_2+T_3=0$ to first order. The invariant J is
\be
J=Im(T_3T_2^*)=Im(T_1T_3^*)=Im(T_2T_1^*)
\ee
which again yields Eq.\ref{J}. Unlike the previous case, the sides of this triangle are sensitive to
the atmospheric parameter $a$ at first order.
One may rescale the sides by $T_1$
\bea
T_1' & = &  1\nonumber \\
T_2' & = & \frac{U_{\mu 2}U_{e2}^*}{U_{\mu 1}U_{e1}^*}\approx -1+\frac{3}{2}re^{i\delta}\nonumber \\
T_3' & = & \frac{U_{\mu 3}U_{e3}^*}{U_{\mu 1}U_{e1}^*}\approx -\frac{3}{2}re^{i\delta}.
\label{Tp}
\eea
As in the previous case,
to first order the rescaled triangle is only sensitive to the reactor parameter $r$ and the phase $\delta$,
which is the anticipated result. To second order
the solar parameter $s$ and the atmospheric parameter $a$ appear.

We now turn to the application of the parametrization
in Eq.\ref{rsa} to neutrino oscillations.
Let us denote by $P_{\alpha \beta}=P(\nu_{\alpha}\rightarrow \nu_{\beta})$ the probability
of transition from a neutrino flavour $\alpha$ to a neutrino flavour $\beta$.
Then expanding to second order in
the parameters $r,s,a$ and $\Delta_{21}$, where it is assumed that
$\Delta_{21} \ll 1$ as in \cite{Akhmedov:2004ny}, we find
considerably simplified vacuum oscillation
\footnote{Similar considerations apply to
oscillations in the presence of matter as discussed in Appendix \ref{matter}.}
probabilities.

The electron anti-neutrino disappearance probability relevant for a
reactor experiment \cite{Goodman:2005ze}
is given
to second order in $r,s,a$ and $\Delta_{21}$ as
\be
P_{ee} = 1-2r^2\sin ^2\Delta_{31} - \frac{8}{9}\Delta_{21}^2
\ee
where $\Delta_{ij}=1.27\Delta m_{ij}^2L/E$ with $L$ the oscillation length in km,
$E$ the beam energy in GeV, and $\Delta m_{ij}^2=m_i^2-m_j^2$ in eV$^2$.
Note that this disappearance probability
is independent of the solar and atmospheric parameters $s,a$, as well
as the phase $\delta$, to this order.

The electron neutrino appearance probability relevant for a forthcoming long baseline muon neutrino
beam experiment \cite{Hayato:2005aj}
is given to second order in $r,s,a$ and $\Delta_{21}$ as
\bea
P_{\mu e } & = & r^2\sin ^2\Delta_{31}+\frac{4}{9}\Delta_{21}^2
+\frac{4}{3}r\Delta_{21}\sin \Delta_{31}\cos (\Delta_{31} + \delta ).
\eea
It is also independent of the solar and atmospheric parameters $s,a$
and only depends on the reactor parameter $r$ and the phase $\delta$ to this order.
The reason is that each of the terms is second order in the parameters $r, \Delta_{21}$,
so any deviations from tri-bimaximal solar or atmospheric mixing only appear
at third order. The muon neutrino disappearance probability
is given to second order in $r,s,a$ and $\Delta_{21}$ as
\bea
P_{\mu \mu} & = & 1-(1-4a^2)\sin ^2\Delta_{31}
- \frac{2}{9}(1+3\cos 2\Delta_{31})\Delta_{21}^2 \nonumber \\
& + &   \frac{2}{3}(1-s-r\cos \delta)\Delta_{21} \sin 2\Delta_{31}.
\eea
Muon neutrino disappearance is clearly sensitive to deviations from tri-bimaximal
mixing, since all three parameters $r,s,a$ and the phase $\delta$ appear.
For example the prospects for measuring deviations from maximal atmospheric mixing
in the next generation of long baseline muon neutrino beam experiments has recently been discussed
\cite{Antusch:2004yx}.
Similarly the tau neutrino appearance probability is given to second order in $r,s,a$ and $\Delta_{21}$ as
\bea
P_{\mu \tau} & = & (1-4a^2-r^2)\sin ^2\Delta_{31} -
\frac{2}{9}(1-3\cos 2\Delta_{31})\Delta_{21}^2
\nonumber \\
& - & \frac{2}{3}(1-s)\Delta_{21} \sin 2\Delta_{31} +  \frac{4}{3}r\Delta_{21}\sin ^2\Delta_{31}\sin \delta.
\eea

We emphasize that the parametrization discussed here is completely
general and is not based on the ansatz of tri-bimaximal mixing, any
more than the Wolfenstein parametrization \cite{Wolfenstein:1983yz} is
based on the ansatz that the quark mixing matrix is equal to the unit
matrix.  Just as the Wolfenstein parametrization is an expansion about
the unit matrix, so this parametrization is an expansion about the
tri-bimaximal matrix.  Unlike the Wolfenstein parametrization, there
are three small parameters $r,s,a$ parametrizing the reactor, solar
and atmospheric deviations from tri-bimaximal mixing.  The expansion
works since the deviations from tri-bimaximal mixing are empirically
small parameters with $r,s,a$ all having magnitude of order the
Wolfenstein parameter $\lambda \approx 0.227$ or less. Indeed these
parameters are sufficiently small that the first order approximation
is accurate enough for many purposes, resulting in quite a simple
looking lepton mixing matrix in Eq.\ref{MNS1}, for example. Unitarity
triangles and neutrino oscillation formulae also have a very simple
form when expressed in this parametrization.

The three parameters $r,s,a$ are not determined at the present time,
and it is even possible that one or more of them (possibly all of
them) are zero, although this seems {\it a priori} unlikely. However,
as mentioned, many speculations appear in the literature as to the
origin and nature of tri-bimaximal mixing and the deviations from it,
and these speculations naturally find expression in this
parametrization. For example certain classes of unified flavour models
\cite{King:2005bj} predict a sum rule which relates $s$ to $r$ and
$\delta$, namely $s\approx r\cos \delta$, where $r\approx \lambda /3$
and $a = O(\lambda^2)$.  Alternatively it has been suggested
\cite{Bjorken:2005rm} that trimaximal solar mixing is exact, $s=0$,
with $a\approx -\frac{1}{2}r\cos \delta$ and $r$ unspecified.  Clearly
an important goal of the next generation of neutrino experiments must
be to show that the parameters $r,s,a$ differ from zero.  Subsequent
high precision neutrino experiments will then be required to
accurately measure the values of the parameters $r,s,a$, as well as
$\delta$, to investigate their possible relationships to each other
and to the Wolfenstein parameter $\lambda$.

\subsection*{Acknowledgments}
We would like to thank J.~Flynn for useful discussions.
We acknowledge partial support from the following grants:
PPARC Rolling Grant PPA/G/S/2003/00096;
EU Network MRTN-CT-2004-503369;
EU ILIAS RII3-CT-2004-506222.

\section*{Appendix}
\appendix
\section{Second order corrections}\label{second}
In this appendix we list the second order corrections to all the results given in the main text.
The second order corrections to the first order MNS matrix elements in Eq.\ref{MNS1} are
\begin{eqnarray}
\Delta U_{e1} & \approx & \sqrt{\frac{2}{3}}(-\frac{1}{4}r^2 -\frac{3}{8}s^2) \nonumber \\
\Delta U_{e2} & \approx & \frac{1}{\sqrt{3}}( -\frac{1}{4}r^2) \nonumber \\
\Delta U_{e3} & \approx & 0\nonumber \\
\Delta U_{\mu 1} & \approx & -\frac{1}{\sqrt{6}}(\frac{1}{2}rse^{i\delta }  -rae^{i\delta }  +sa +a^2) \nonumber \\
\Delta U_{\mu 2} & \approx & \frac{1}{\sqrt{3}}(-\frac{1}{2}rse^{i\delta }  -\frac{1}{2}rae^{i\delta }  +\frac{1}{2}sa - \frac{3}{8}s^2-a^2)
\nonumber \\
\Delta U_{\mu 3} & \approx & \frac{1}{\sqrt{2}}(-\frac{1}{4}r^2)
\nonumber \\
\Delta U_{\tau 1} & \approx &
\frac{1}{\sqrt{6}}(\frac{1}{2}rse^{i\delta }  +rae^{i\delta }  +sa)
\nonumber \\
\Delta U_{\tau 2} & \approx &
-\frac{1}{\sqrt{3}}(\frac{1}{2}rse^{i\delta }  -\frac{1}{2}rae^{i\delta }  -\frac{1}{2}sa - \frac{3}{8}s^2)
\nonumber \\
\Delta U_{\tau 3} & \approx &
\frac{1}{\sqrt{2}}(-\frac{1}{4}r^2 -a^2).
\label{MNS2}
\end{eqnarray}

The second order correction to the Jarlskog CP invariant in Eq.\ref{J} is
\be
\Delta J\approx  \frac{rs}{12} \sin \delta .
\ee

The second order corrections to the unscaled sides of the $\nu_2.\nu_3$ unitarity
triangle in Eq.\ref{S} are
\bea
\Delta S_1 & \approx & \frac{1}{\sqrt{6}}sre^{i\delta} \nonumber \\
\Delta S_2 & \approx &  -\frac{1}{\sqrt{6}}(\frac{r^2}{4} +2a^2 +\frac{3}{8}s^2+are^{i\delta}+\frac{1}{2}sre^{i\delta} )\nonumber \\
\Delta S_3 & \approx &  \frac{1}{\sqrt{6}}(\frac{r^2}{4} +2a^2 +\frac{3}{8}s^2+are^{i\delta}-\frac{1}{2}sre^{i\delta}).
\eea
The second order corrections to the normalized sides of the $\nu_2.\nu_3$ unitarity
triangle in Eq.\ref{Sp} are
\bea
\Delta S_1' & \approx & \frac{r^2}{2}e^{2i\delta} -\frac{3}{2}sre^{i\delta}\nonumber \\
\Delta S_2' & \approx & -\frac{r^2}{2}e^{2i\delta} +\frac{3}{2}sre^{i\delta}\nonumber \\
\Delta S_3' & = &  0.
\eea

The second order corrections to the unscaled sides of the $\nu_e.\nu_{\mu}$ unitarity
triangle in Eq.\ref{T} are
\bea
\Delta T_1 & \approx & \frac{r^2}{12}+\frac{7}{24}s^2+\frac{a^2}{3}+\frac{sa}{6} +\frac{sr}{3}e^{i\delta}
-\frac{ar}{3}e^{i\delta} \nonumber \\
\Delta T_2 & \approx &
-\frac{r^2}{12}-\frac{7}{24}s^2-\frac{a^2}{3}-\frac{sa}{6} -\frac{sr}{3}e^{i\delta}
-\frac{ar}{6}e^{i\delta}
\nonumber \\
\Delta T_3 & \approx & \frac{ar}{2}e^{i\delta}.
\eea
The second order corrections to the normalized sides of the $\nu_e.\nu_{\mu}$ unitarity
triangle in Eq.\ref{Tp} are
\bea
\Delta T_1' & = & 0\nonumber \\
\Delta T_2' & \approx & -\frac{3}{4}re^{i\delta}(2re^{i\delta}+s-4a)\nonumber \\
\Delta T_3' & \approx &  \frac{3}{4}re^{i\delta}(2re^{i\delta}+s-4a).
\eea

\section{Neutrino oscillations in matter}\label{matter}
In this appendix we present the complete formulae for neutrino oscillations in the
presence of matter of constant density to second order in the quantities
$r,s,a$ and $\Delta_{21}$, where it is assumed that
$\Delta_{21} \ll 1$ as in \cite{Akhmedov:2004ny}.
Following \cite{Akhmedov:2004ny} we write
$\Delta = \Delta_{31}$, $\alpha = \frac{\Delta m_{21}^2}{\Delta m_{31}^2}$
and $A=\frac{VL}{2\Delta}$ where $V$ is the potential expressed in units of eV as
\be
V\approx 7.56\times 10^{-14}\rho \ Y_e
\ee
where $\rho$ is the matter density of the Earth in units of g/cm$^3$ and
$Y_e\approx 0.5$ is the number of electrons per nucleon in the Earth.
The constant density approximation is good when the neutrino beam only passes
through the Earth's crust where $\rho \approx 3$ g/cm$^3$ or the
Earth's mantle where $\rho \approx 4.5$ g/cm$^3$.

The complete set of neutrino oscillation probabilities for electron neutrino
or muon neutrino beams in the
presence of matter of constant density to second order in the parameters
$r,s,a$ and $\alpha$ are
\beq
P_{ee}  =  1 - \frac{8}{9}\alpha^2 \frac{\sin ^2 A\Delta }{A^2}
-2r^2\frac{\sin ^2 (A-1)\Delta }{(A-1)^2}.
\ee
\bea
P_{e\mu} & = & \frac{4}{9}\alpha^2 \frac{\sin ^2 A\Delta }{A^2}
+ r^2\frac{\sin ^2 (A-1)\Delta }{(A-1)^2} \nonumber \\
& + &
\frac{4}{3}r\alpha  \cos (\Delta - \delta ) \frac{\sin A\Delta }{A} \frac{\sin (A-1)\Delta }{(A-1)}.
\eea
\bea
P_{e\tau} & = &
\frac{4}{9}\alpha^2 \frac{\sin ^2 A\Delta }{A^2}
+ r^2\frac{\sin ^2 (A-1)\Delta }{(A-1)^2} \nonumber \\
& - &
\frac{4}{3}r\alpha  \cos (\Delta - \delta ) \frac{\sin A\Delta }{A} \frac{\sin (A-1)\Delta }{(A-1)}
\eea
\bea
P_{\mu e} & = & \frac{4}{9}\alpha^2 \frac{\sin ^2 A\Delta }{A^2}
+ r^2\frac{\sin ^2 (A-1)\Delta }{(A-1)^2} \nonumber \\
& + &
\frac{4}{3}r\alpha  \cos (\Delta + \delta ) \frac{\sin A\Delta }{A} \frac{\sin (A-1)\Delta }{(A-1)}.
\eea
\bea
P_{\mu \mu} & = & 1
- (1-4a^2)\sin ^2 \Delta
+ \frac{2}{3}(1-s)\alpha \Delta \sin 2 \Delta \nonumber \\
& - & \frac{4}{9} \alpha^2 \frac{\sin ^2 A\Delta }{A^2}
-\frac{4}{9}\alpha^2 \Delta^2 \cos 2\Delta \nonumber \\
& + &
\frac{4}{9}\alpha^2 \frac{1}{A}
\left( \sin \Delta \frac{\sin A\Delta }{A} \cos (A-1)\Delta - \frac{\Delta }{2}\sin 2\Delta  \right)
\nonumber \\
& - & r^2\frac{\sin ^2 (A-1)\Delta }{(A-1)^2} \nonumber \\
& - & \frac{1}{A-1}r^2
\left( \sin \Delta \cos A\Delta \frac{\sin (A-1)\Delta }{(A-1)} - \frac{A}{2}\Delta \sin 2\Delta  \right)
           \nonumber \\
& - &
\frac{4}{3}r\alpha  \cos \delta \cos \Delta \frac{\sin A\Delta }{A} \frac{\sin (A-1)\Delta }{(A-1)}.
\eea
\bea
P_{\mu \tau} & = &
(1-4a^2)\sin ^2 \Delta
- \frac{2}{3}(1-s)\alpha \Delta \sin 2 \Delta
+ \frac{4}{9}\alpha^2 \Delta^2 \cos 2\Delta   \nonumber \\
& - &
\frac{4}{9}\alpha^2 \frac{1}{A}
\left( \sin \Delta \frac{\sin A\Delta }{A} \cos (A-1)\Delta - \frac{\Delta }{2}\sin 2\Delta  \right)
\nonumber \\
& + &
\frac{1}{A-1}r^2
\left( \sin \Delta \cos A\Delta \frac{\sin (A-1)\Delta }{(A-1)} - \frac{A}{2}\Delta \sin 2\Delta  \right)
           \nonumber \\
& + &
\frac{4}{3}r\alpha  \sin \delta \sin \Delta \frac{\sin A\Delta }{A}
\frac{\sin (A-1)\Delta }{(A-1)}.
\eea

\end{document}